\documentclass[prl,twocolumn,showpacs,floatfix,,amsmath,amssymb,superscriptaddress]{revtex4}
\usepackage{amsfonts}
\usepackage{stmaryrd}
\usepackage{bbm}
\usepackage{mathrsfs}
\usepackage{tipa}
\usepackage{amssymb}
\usepackage{txfonts}
\usepackage{graphicx}
\usepackage{dcolumn}
\usepackage{epstopdf}
\usepackage[colorlinks,linkcolor=blue,urlcolor=blue,citecolor=blue]{hyperref}
\usepackage{multirow}
\usepackage{subfigure}

\begin{document}
\newcommand*{\cm}{cm$^{-1}$\,}
\newcommand*{\Tc}{T$_c$\,}

\newcommand {\C}[1]{\textcolor{red}{#1}}
\newcommand {\B}[1]{\textcolor{blue}{#1}}



\title{
Magneto-infrared spectroscopy of
Landau levels and Zeeman splitting of three-dimensional massless Dirac Fermions in ZrTe$_5$
}

\author{R. Y. Chen$^{\dagger}$}
\affiliation{International Center for Quantum Materials, School of Physics, Peking University, Beijing 100871, China}

\author{Z. G. Chen$^{\dagger}$}
\affiliation{National High Magnetic Field Laboratory, Tallahassee, Florida 32310, USA}

\author{X.-Y. Song}
\affiliation{International Center for Quantum Materials, School of Physics, Peking University, Beijing 100871, China}

\author{J. A. Schneeloch}
\affiliation{Condensed Matter Physics and Materials Science
Department, Brookhaven National Lab, Upton, New York 11973, USA}

\author{G. D. Gu}
\affiliation{Condensed Matter Physics and Materials Science
Department, Brookhaven National Lab, Upton, New York 11973, USA}

\author{F. Wang}
\email{wangfa@pku.edu.cn}
\affiliation{International Center for Quantum Materials, School of Physics, Peking University, Beijing 100871, China}
\affiliation{Collaborative Innovation Center of Quantum Matter, Beijing, China}

\author{N. L. Wang}
\email{nlwang@pku.edu.cn}
\affiliation{International Center for Quantum Materials, School of Physics, Peking University, Beijing 100871, China}
\affiliation{Collaborative Innovation Center of Quantum Matter, Beijing, China}

\begin{abstract}
We present a magneto-infrared spectroscopy study on a newly identified three-dimensional (3D) Dirac semimetal ZrTe$_5$. We observe clear transitions between Landau levels and their further splitting under magnetic field.
Both the sequence of transitions and their field dependence follow quantitatively the relation expected for 3D \emph{massless} Dirac fermions. The measurement also
reveals an exceptionally low magnetic field needed to drive the compound into its quantum limit, demonstrating that ZrTe$_5$ is an extremely clean system and ideal platform for studying 3D Dirac fermions. The splitting of the Landau levels provides a direct and bulk spectroscopic evidence that a relatively weak magnetic field can produce a sizeable Zeeman effect on the 3D Dirac fermions, which lifts the spin degeneracy of Landau levels. Our analysis indicates that the compound evolves from a Dirac semimetal into a topological line-node semimetal under current magnetic field configuration.

\end{abstract}


\maketitle

3D topological Dirac/Weyl semimetals are new kinds of topological materials
that possess linear band dispersion in the bulk along all three momentum directions\cite{PhysRevB.85.195320,PhysRevB.88.125427,PhysRevB.83.205101,Liu2014-Science,Liu2014-NatMater,Borisenko2014,Neupane2014}.
Their low-energy quasiparticles are the condensed matter realization of Dirac and Weyl fermions in relativistic
high energy physics\cite{Young2012,Kharzeev2014}. These materials are expected to host many unusual phenomena\cite{He2014a,Liang2015c,Xu2015-Science},
in particular the chiral and axial anomaly associated with Weyl fermions\cite{PhysRevB.83.205101,Chen2013a,Hosur2014,Potter2014}.
It is well known that the Dirac nodes are protected by both time-reversal and space inversion symmetry.
Since magnetic field breaks the time-reversal symmetry, a Dirac node may be split into a pair of Weyl nodes along the magnetic field direction in the momentum space \cite{PhysRevLett.107.127205,PhysRevB.84.235126,Gorbar-MagneticDiracWeyl} or transformed
into line-nodes\cite{PhysRevB.84.235126,PhysRevB.90.115111}.
Therefore, a Dirac semimetal
can be considered as a parent compound to realize other topological variant quantum states. However, past 3D Dirac semimetal materials (e.g. Cd$_3$As$_2$) suffer from the problem of large residual carrier density which requires very high magnetic field (e.g. above 60 Tesla) to drive them to their quantum limit\cite{PhysRevLett.114.117201,JCao}. This makes it extremely difficult to explore the transformation from Dirac to Weyl or line-node semimetals. Up to now, there are no direct evidences of such transformations.

ZrTe$_5$ appears to be a new topological 3D Dirac material that exhibits novel and interesting properties. The compound crystallizes in the layered orthorhombic crystal structure,
with prismatic ZrTe$_6$ chains running along the crystallographic $a$-axis
and linked along the $c$-axis via zigzag chains of Te atoms to form two-dimensional (2D) layers.
Those layers stack along the $b$-axis.
A recent \emph{ab initio} calculation suggests that bulk ZrTe$_5$ locates close to
the phase boundary between weak and strong topological insulators \cite{PhysRevX.4.011002}.
However, more recent transport and ARPES experiments identify it to be a 3D Dirac semimetal with only one Dirac node at the $\Gamma$ point \cite{LiQ}.
Interestingly, a chiral magnetic effect associated with the transformation from a Dirac to Weyl semimetal was observed on ZrTe$_5$ through a magneto-transport measurement \cite{LiQ}.
Our recent optical spectroscopy measurement at zero field revealed clearly a linear energy dependence of optical conductivity, being another hallmark of 3D massless Dirac fermions \cite{RYChen}.

In this letter, we present magneto-infrared spectroscopy study on ZrTe$_5$ single crystals. We observe clear transitions between Landau levels and their further splitting under magnetic field.
Both the sequence of transitions and their field dependence follow quantitatively the relation expected for 3D \emph{massless} Dirac fermions. Furthermore, the measurement
reveals an exceptionally low magnetic field (about 1 Tesla) needed to drive the compound into its quantum limit. Both facts demonstrate that ZrTe$_5$ is an extremely clean system and ideal platform for studying 3D Dirac fermions. The presence of further splitting of Landau levels, which has never been observed in 2D massless Dirac fermions, e.g. graphene, provides direct evidence for the lifting of spin degeneracy of Landau levels, an effect being linked to the transformation from a Dirac semimetal to a line-node or Weyl semimetal. Our theoretical analysis indicates that the former one is more likely realized in the present magnetic field configuration.

\begin{figure}
\centering
\includegraphics[width=9cm]{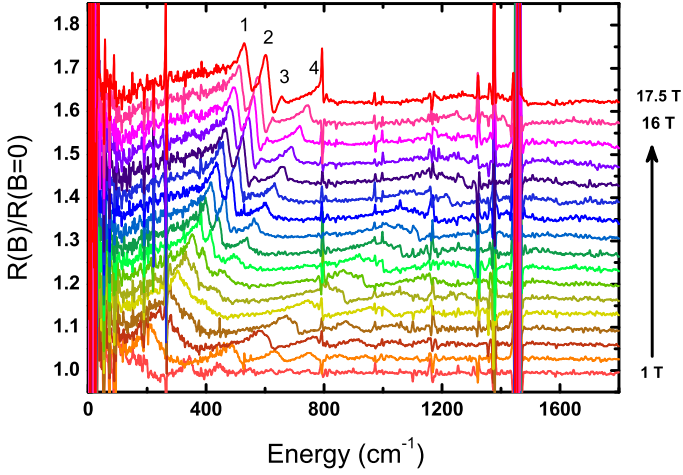}\\
\caption{
The relative reflectivity of ZrTe$_5$ under magnetic field, as a function of energy.
The spectrum are shifted upward by equal interval corresponding to different $B$.}
\label{Fig:RB}
\end{figure}

Figure~\ref{Fig:RB} shows the reflectance spectra under different magnetic field $R(B)$ renormalized by the zero field reflectance $R(0)$ in the far- and mid-infrared region.
For the lowest magnetic field (1 $T$), a series of peaks could be clearly resolved, which keep growing more pronounced and shift to higher energies when the field strength $B$ increases.
In optical reflectance measurement, such peak features usually come from the interband transitions.
Since these sharp peaks emerge in the reflectivity only by applying magnetic field,
it is natural to connect them to the Landau quantization of 3D Dirac electrons.
Thus the peaks should stem from electronic transitions connecting different Landau levels.
Significantly, the first broad peak, which appears at the lowest energy, gradually split into four narrow peaks as $B$ increases.
This character is quite intriguing and has never been observed ever before,
which will be explained in detail later.
For the sake of convenience, they are marked by the numbers ``1, 2, 3, 4'' respectively at the top of Fig.~\ref{Fig:RB}.
However, the splittings of other peaks located at higher energies are rather vague in this plot.

In order to verify our speculation on the origin of the emerging peaks
and capture the underlying physics of 3D Dirac semimetal,
we examine the sequence of peaks observed at low field.
For a 3D system, the band structure would transform into a set of 1D Landau levels by applying strong enough magnetic field, which are only dispersive along the field direction.
Theoretical calculation on an isolated Weyl point has suggested that the magneto-optical conductivity is constituted of a series of asymmetric peaks lying on top of a linear background \cite{Hosur2012,Ashby2013}.
Especially, the peaks associated with allowed interband transitions in the Landau level structure occur at $\omega\propto \sqrt{n}+\sqrt{n+1}$, corresponding to transition from $L_{-n}$ to $L_{n+1}$ or from $L_{-(n+1)}$ to $L_n$, where $L_n$ represents for the $n$th Landau level.
This conclusion applies to 3D massless Dirac fermion as well,
because massless Dirac fermion can be thought as two sets of Weyl fermions with opposite chirality.

\begin{figure}
\centering
\includegraphics[width=7cm]{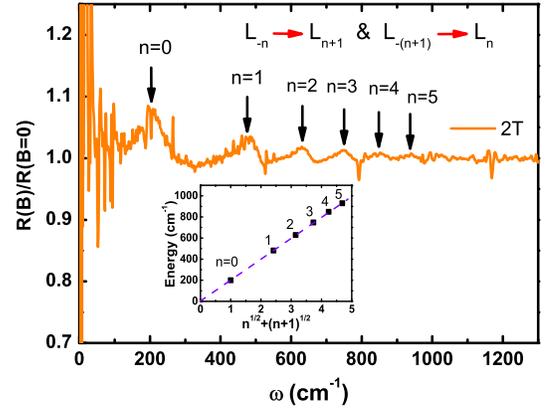}\\
\caption{
The wave number dependent relative reflectivity $R(B)/R(0)$ under magnetic field of 2 T. $n=0\ldots 5$ point to six different peaks in sequence. The inset shows the linear dependence of the transition energies between Landau levels on $\sqrt{n}+\sqrt{n+1}$. The dash line is a guide to the eyes.}
\label{Fig:LL}
\end{figure}

A linear rising optical conductivity has been revealed in our previous zero field spectroscopic experiment on ZrTe$_5$ single crystal \cite{RYChen}
which already provides a strong evidence for 3D massless Dirac or Weyl femions.
We blew up the results of $B=2$ T as displayed in Fig.~\ref{Fig:LL},
in which six peaks could be clearly resolved.
The positions of these peaks are
identified to be about
$202$, $480$, $628$, $748$, $856$, $937$ \cm in sequence.
The energy ratios
of the peaks observed here can be approximately reduce to 1 : 1+$\sqrt2$ : $\sqrt2$+$\sqrt3$ : $\sqrt3$+$\sqrt4$ : $\sqrt4$+$\sqrt5$ : $\sqrt5$+$\sqrt6$, in nearly perfect accordance with the predicted massless Dirac semimetal behavior.
This results in a linear dependence of the transition energy between Landau levels on $\sqrt{n}+\sqrt{n+1}$, as shown in the inset of the figure.
From this observation,
the first peak can be unambiguously determined to correspond to $n$=0, ascribed to transitions from $L_0$ to $L_1$ and $L_{-1}$ to $L_0$.
Only when the chemical potential lies in between $L_1$ and $L_{-1}$, can this transition be clearly visible.\cite{Ashby2013}
This is quite exciting because it demonstrates that the quantum limit could be easily approached by magnetic field as low as $1$ T, where the $n=0$ peak is distinctively observed.
As a contrast, the quantum limit of the well-known Dirac semimetal Cd$_3$As$_2$ can not be reached with magnetic field lower than $65$ T \cite{PhysRevLett.114.117201,JCao}.
It indicates that ZrTe$_5$ compound is extremely close to an ideal Dirac semimetal, with the chemical potential lying in the vicinity of Dirac point, and meanwhile it is an extraordinarily clean system.
We anticipate that our finding of easy access of quantum limit will motivate many other experimental probes on this compound.
In Supplemental Information \cite{SI}
we perform more detailed analysis of this peak sequence
and obtain estimates of the average $ac$-plane Fermi velocity
$v_{\perp}=\sqrt{v_a v_c} \sim 4.84\times 10^5$ (m/s) and
a vanishingly small Dirac mass $|m|\sim 2$ (cm$^{-1}$).

\begin{figure}
\centering
\includegraphics[width=8cm]{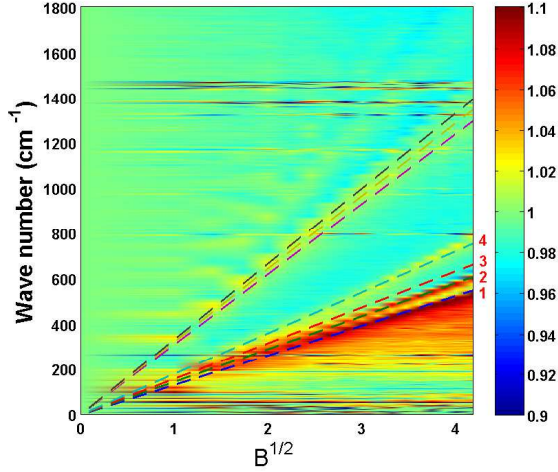}\\
\caption{
The pseudo-colors photograph of the relative reflectivity $R(B)/R(0)$ as functions of wave number and $\sqrt{B}$. The dashed lines are linear fittings of the peak energies dependent on $\sqrt{B}$.}
\label{Fig:mag}
\end{figure}

In Fig.~\ref{Fig:RB}, where $R(B)/R(0)$ was shifted by equal interval with regard to increasing magnetic field, it is noted that the peak positions evolve
in a way much alike the parabolic fashion as $B$. To further illustrate the characteristic features of the Landau levels, we plot $R(B)/R(0)$ in a pseudo-color photograph as a function of $\sqrt{B}$. It is clearly seen in Fig.~\ref{Fig:mag} that the wave numbers of the peaks are basically linear proportional to $\sqrt{B}$. The dashed lines are instructions for eyes, whose intercepts at $0$ T are all absolute zero.
For a single massless 3D Dirac node, the $n$th Landau levels caused by magnetic field
are dispersive only along the field direction
(see  Supplementary Information\cite{SI}
for more details),
with doubly degenerate $n\neq 0$ levels
$E_n(k_{\parallel}) = \text{sgn}(n)\cdot \sqrt{2 v_{\perp}^2 eB\hbar\cdot |n|+\hbar^2 v_{\parallel}^2 k_{\parallel}^2}$,
and
$E_0(k_{\parallel})=\pm \hbar v_{\parallel} k_{\parallel}$ for $n=0$.
If neglecting the dispersion along the magnetic field direction, then $E_n\propto \sqrt{B}$.
As a comparison, for free electron systems, the magnetic induced Landau level obeys $E_n=(n+\frac{1}{2})\hbar\omega_c$, where $\omega_c$ is the the cyclotron angular velocity and proportional to $B$ instead of $\sqrt{B}$. Additionally, the Landau level energy of massive 3D Dirac fermions in
topological insulator Bi$_2$Se$_3$ is reported to be in linear scale with $B$ as well\cite{Bi2Se3}. Therefore, the $\sqrt{B}$ dependence in Fig.~\ref{Fig:mag} intensively implies again the characteristic property of 3D massless Dirac fermions in ZrTe$_5$ under magnetic field.

In addition to the sequence of peaks at low field, the peak splitting shown in Fig.~\ref{Fig:RB} could also be well resolved in Fig.~\ref{Fig:mag}. The $n=0$ peak evolves into four peaks at high magnetic field, with two very pronounced ones at lower energies and two relatively weak ones at higher energies. They are labeled as "1, 2, 3, 4" respectively in accordance with Fig.~\ref{Fig:RB} The splittings of the rest peaks are, although too vague to be precisely identified, but for certain to exist. The $n=1$ peak, arising from transitions from $L_{-2}$ to $L_1$ and $L_{-1}$ to $L_2$, seems to split into 3 peaks. Such splitting has never been observed in 2D massless Dirac fermion system, for example graphene\cite{PhysRevLett.97.266405}.

\begin{figure}
\centering
\includegraphics[width=8cm]{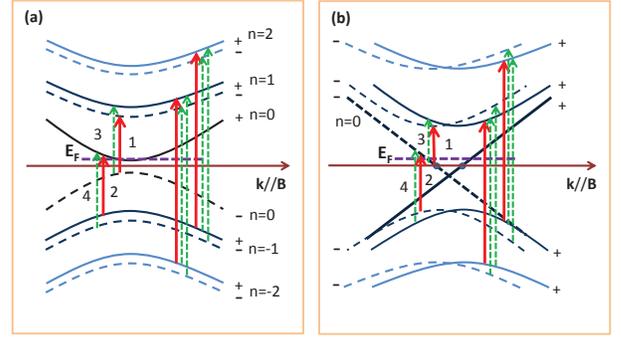}\\
\caption{
Two proposed scenarios. The left panel (a) shows the splitting of Landau levels simply caused by Zeeman effect. The strong spin-orbit coupling could lead to a mixing of spin up and down components. So the split "+" (solid lines) and "-" (dashed lines) levels do not have pure spin up and spin down components. This would allow inter Landau level transitions to occur with opposite signs, however, weak in intensity. The purple dashed line is the Fermi energy $E_F$.
Zeeman field without orbital effect will transform
3D Dirac node into ``line-nodes'' in this case\cite{PhysRevB.84.235126}.
The right panel
(b) represents for the Landau levels of a Weyl semimetal induced by the magnetic field.
The crossing points of $n=0$ Landau levels with the horizontal axis
are the would-be Weyl points, if the orbital effect of magnetic field is ignored.
The solid lines and dashed lines represent
the two sets of Landau levels from the two Weyl nodes with
opposite chirality.
Transitions between Landau levels of opposite chirality
will have weaker intensity.
In both panels, the red solid arrows represent for transitions between Landau levels of the same spin/chirality, whereas the green dashed ones indicate that of different spin/chirality.
}
\label{Fig:zeeman}
\end{figure}

We now explore the underlying mechanism for the splitting.
We will show that this splitting can be naturally explained by
the Zeeman effects of magnetic field on 3D Dirac fermions.
The effect of Zeeman field on Dirac semimetals
has been thoroughly studied by Burkov \emph{et al.}\cite{PhysRevB.84.235126}.
It was pointed out that
Zeeman field may split the Dirac node
into two Weyl nodes, or
transform the Dirac node into ``line-nodes''\cite{PhysRevB.84.235126}.
Further including orbital effects of magnetic field
will generate Zeeman-split Landau levels,
schematically shown in Fig.~\ref{Fig:zeeman}.
Here we will briefly explain the two possible scenarios
and leave the details in the
Supplementary Information\cite{SI}.

The first scenario we consider is the ``line-nodes'' picture,
with the resulting Landau level structure depicted in Fig.~\ref{Fig:zeeman}(a).
The doubly degenerate Landau levels $E_n(k_{\parallel})$ for $n\neq 0$ from
3D Dirac fermions
will be split into $E_{n,\pm}\sim E_n(k_{\parallel})\pm \bar{g}\mu_B B/2$,
where $\bar{g}$ is the average Land\'e $g$-factor
for the conduction and valence bands of 3D Dirac fermion.
The $n=0$ Landau levels $E_0(k_{\parallel})=\pm \hbar v_{\parallel} k_{\parallel}$
will mix around $k_{\parallel}=0$ and open a gap of size $\bar{g}\mu_B B$ there,
and become $E_{0,\pm}\sim \pm \sqrt{\hbar^2 v_{\parallel}^2 k_{\parallel}^2+(\bar{g}\mu_B B/2)^2}$.
The split Landau levels are labeled by ``spin'' indices `$+$' and `$-$'
in Fig.~\ref{Fig:zeeman}(a),
which indicate that the states are of spin up or spin down at $k_{\parallel}=0$.
However with $k_{\parallel}\neq 0$ the split Landau levels
are not purely spin up or spin down due to strong spin-orbit coupling.
Optical transitions between levels of the same ``spin'' indices
can happen at $k_{\parallel}=0$, and produce strong peaks in optical conductivity
(thus reflectivity)\cite{Ashby2013},
but transitions between levels of opposite spin indices
will be suppressed at $k_{\parallel}=0$, leading to weak and broad peaks.
Therefore the original $n=0$ peak will split into two strong peaks ``1'' from
$L_{0,-}$ to $L_{1,-}$, and ``2'' from $L_{-1,+}$ to $L_{0,+}$,
and two weak peaks ``3'' from
$L_{0,-}$ to $L_{1,+}$, and ``4'' from $L_{-1,-}$ to $L_{0,+}$.
The peak ``1'' and peak ``2'' can have different energy
if the chemical potential is not in the gap between $L_{0,-}$ and $L_{0,+}$
as depicted in Fig.~\ref{Fig:zeeman}(a),
or if the conduction and valence bands of 3D Dirac fermion
have different $g$-factor (see
Supplementary Information\cite{SI}).
In any case
the splitting between peak ``3'' and peak ``1'',
and between peak ``4'' and peak ``2''
will be about $\bar{g}\mu_B B$.
Based on previous \emph{ab initio} results\cite{PhysRevX.4.011002} on ZrTe$_5$ and the experimental setup
we conclude that this scenario is the most likely explanation of our observation (see
Supplementary Information\cite{SI}
for details).

The second and more interesting scenario is the ``Weyl nodes'' picture
illustrated in Fig.~\ref{Fig:zeeman}(b).
In this case the Zeeman field effectively shifts
the wave vector parallel to field by
$\pm \bar{g}\mu_B B/2\hbar v_{\parallel}$,
where the $\pm$ sign depends on the chirality of Weyl fermions.
The degenerate $n\neq 0$ Landau levels become
$E_{n,\pm}(k_{\parallel})\sim E_n(k_{\parallel}\mp \bar{g}\mu_B B/2\hbar v_{\parallel})$,
and the $n=0$ Landau levels
become
$E_{0,\pm}(k_{\parallel})\sim \pm \hbar v_{\parallel} k_{\parallel}-\bar{g}\mu_B B/2$.
In this scenario transitions between Landau levels of the same (different) chirality
will have strong (weak) intensity.
The original $n=0$ peak will also split into two strong peaks ``1'' from
$L_{0,-s}$ to $L_{1,-s}$, and ``2'' from $L_{-1,-s}$ to $L_{0,-s}$,
and two weak peaks ``3'' from
$L_{0,-s}$ to $L_{1,s}$, and ``4'' from $L_{-1,s}$ to $L_{0,-s}$,
where $s=\pm$.
The peak ``1'' and peak ``2'' can have different energy
if the chemical potential is not at the charge neutrality
as depicted in Fig.~\ref{Fig:zeeman}(b).
However the splitting between peak ``3'' and ``1'' will in general
not be linear in $B$,
unless the conduction and valence bands have very different $g$-factors.
According to our analysis\cite{SI}
this scenario is more likely to happen when the field is applied
along crystal $c$-direction.

\begin{figure}
\centering
\includegraphics[width=6cm]{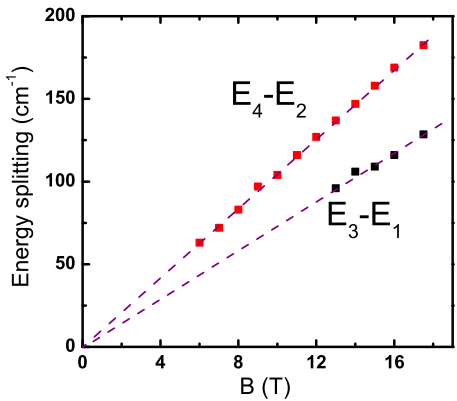}\\
\caption{
The magnetic field dependent energy difference of $E_3$-$E_1$ and $E_4$-$E_2$.
They represent the Landau level splittings for the conduction and valence bands, respectively.
}
\label{Fig:linear}
\end{figure}

From the peak positions of ``1'', ``2'', ``3'' and ``4'', we can immediately obtain the dependence of the split energy as a function of magnetic field. Figure~\ref{Fig:linear} displays the magnetic field dependence of $E_4-E_2$ above 6 T and $E_3$-$E_1$ above $13$ T, respectively. The energy positions for peak "3" could not be well resolved below 12 T, so the energy difference of $E_3-E_1$ is plotted only at high magnetic field. Obviously, both $E_4-E_2$ and $E_3$-$E_1$ exhibit good linear dependence, in better
agreement with
the first ``line-nodes'' scenario.
The energy splitting is roughly 10.5 cm$^{-1}$/T ($\sim$1.3 meV/T) for $E_4-E_2$ and 7.4 cm$^{-1}$/T ($\sim$0.92 meV/T) for $E_3-E_1$.
According to previous discussions,
this leads to estimates of the average $g$-factor being $22.5$ or $15.8$,
from the $E_4 - E_2$ or $E_3 - E_1$ splittings respectively.

It was known that the $g$-factor reaches about 37 for Cd$_3$As$_2$\cite{Jeon2014}, which is even bigger than our estimated $g$-factor for
the ZrTe$_5$ compound.
We do not yet have a good explanation for the discrepancy between $E_4 - E_2$ and $E_3 - E_1$ splittings.
Based on our analysis in Supplementary Information\cite{SI},
this could be due to the broad nature of the weaker peaks,
and the current analysis may overestimate the $g$-factors
for this reason.

The transitions associated with $L_{-2}\rightarrow L_{1}$ and $L_{-1}\rightarrow L_2$ are much more complex. Analogy to the transitions between $L_0$ and $L_{\pm 1}$, the
selection rule permissive ones (indicated by red solid arrows) should be more pronounced than those between opposite spin orientations (the dashed green ones). Considering possible different splittings of the valence and conduction bands,
the $n=1$ peak appeared in $R(B)/R(0)$ is supposed to contains 
at least three components.

In summary, by performing magneto-optical measurement on the single crystalline 3D massless Dirac semimetal ZrTe$_5$, we have clearly observed the magnetic field induced Landau levels, evidenced by regular organized peaks shown in the renormalized reflectivity $R(B)/R(0)$.
Particularly, the first peak is identified to be originated from the transitions between the zeroth and first Landau Levels, which reveals the Fermi energy lies very close to the Dirac point.
The appearance of the first peak under magnetic field as low as $1$ T demonstrates an exceptionally low quantum limit of ZrTe$_5$ compared to other 3D Dirac semimetals, which provided an elegant platform to explore more intriguing non-trivial quantum phenomena.
Of most importance, the fourfold splitting of the first peak yield direct and clear evidence for the release of spin degeneracy of Landau level, hence the transformation from a Dirac semimetal into line-node or Weyl semimetal. Our theoretical modeling and analysis indicate that the former one is more likely realized in the present magnetic field configuration.

\begin{center}
\small{\textbf{ACKNOWLEDGMENTS}}
\end{center}

$^{\dagger}$ These authors contributed equally to this work. We acknowledge very helpful discussions with Z. Fang, H. M. Weng, X. C. Xie, D. H. Lee, L. Fu, Q. Li, X. Dai, H. W. Liu. This work was supported by the National Science Foundation of China
(11120101003, 11327806, 11374018), and
the 973 project of the Ministry of Science and Technology of China
(2011CB921701, 2012CB821403, 2014CB920902).
Work at Brookhaven is supported by the Office of Basic Energy Sciences, Division of Materials Sciences and Engineering, U.S. Department of Energy under Contract No. DE-SC00112704.

\bibliography{ZrTe5}

\begin{thebibliography}{30}
\expandafter\ifx\csname natexlab\endcsname\relax\def\natexlab#1{#1}\fi
\expandafter\ifx\csname bibnamefont\endcsname\relax
  \def\bibnamefont#1{#1}\fi
\expandafter\ifx\csname bibfnamefont\endcsname\relax
  \def\bibfnamefont#1{#1}\fi
\expandafter\ifx\csname citenamefont\endcsname\relax
  \def\citenamefont#1{#1}\fi
\expandafter\ifx\csname url\endcsname\relax
  \def\url#1{\texttt{#1}}\fi
\expandafter\ifx\csname urlprefix\endcsname\relax\def\urlprefix{URL }\fi
\providecommand{\bibinfo}[2]{#2}
\providecommand{\eprint}[2][]{\url{#2}}

\bibitem[{\citenamefont{Wang et~al.}(2012)\citenamefont{Wang, Sun, Chen,
  Franchini, Xu, Weng, Dai, and Fang}}]{PhysRevB.85.195320}
\bibinfo{author}{\bibfnamefont{Z.}~\bibnamefont{Wang}},
  \bibinfo{author}{\bibfnamefont{Y.}~\bibnamefont{Sun}},
  \bibinfo{author}{\bibfnamefont{X.-Q.} \bibnamefont{Chen}},
  \bibinfo{author}{\bibfnamefont{C.}~\bibnamefont{Franchini}},
  \bibinfo{author}{\bibfnamefont{G.}~\bibnamefont{Xu}},
  \bibinfo{author}{\bibfnamefont{H.}~\bibnamefont{Weng}},
  \bibinfo{author}{\bibfnamefont{X.}~\bibnamefont{Dai}}, \bibnamefont{and}
  \bibinfo{author}{\bibfnamefont{Z.}~\bibnamefont{Fang}},
  \bibinfo{journal}{Phys. Rev. B} \textbf{\bibinfo{volume}{85}},
  \bibinfo{pages}{195320} (\bibinfo{year}{2012}),
  \urlprefix\url{http://link.aps.org/doi/10.1103/PhysRevB.85.195320}.

\bibitem[{\citenamefont{Wang et~al.}(2013)\citenamefont{Wang, Weng, Wu, Dai,
  and Fang}}]{PhysRevB.88.125427}
\bibinfo{author}{\bibfnamefont{Z.}~\bibnamefont{Wang}},
  \bibinfo{author}{\bibfnamefont{H.}~\bibnamefont{Weng}},
  \bibinfo{author}{\bibfnamefont{Q.}~\bibnamefont{Wu}},
  \bibinfo{author}{\bibfnamefont{X.}~\bibnamefont{Dai}}, \bibnamefont{and}
  \bibinfo{author}{\bibfnamefont{Z.}~\bibnamefont{Fang}},
  \bibinfo{journal}{Phys. Rev. B} \textbf{\bibinfo{volume}{88}},
  \bibinfo{pages}{125427} (\bibinfo{year}{2013}),
  \urlprefix\url{http://link.aps.org/doi/10.1103/PhysRevB.88.125427}.

\bibitem[{\citenamefont{Wan et~al.}(2011)\citenamefont{Wan, Turner, Vishwanath,
  and Savrasov}}]{PhysRevB.83.205101}
\bibinfo{author}{\bibfnamefont{X.}~\bibnamefont{Wan}},
  \bibinfo{author}{\bibfnamefont{A.~M.} \bibnamefont{Turner}},
  \bibinfo{author}{\bibfnamefont{A.}~\bibnamefont{Vishwanath}},
  \bibnamefont{and} \bibinfo{author}{\bibfnamefont{S.~Y.}
  \bibnamefont{Savrasov}}, \bibinfo{journal}{Phys. Rev. B}
  \textbf{\bibinfo{volume}{83}}, \bibinfo{pages}{205101}
  (\bibinfo{year}{2011}),
  \urlprefix\url{http://link.aps.org/doi/10.1103/PhysRevB.83.205101}.

\bibitem[{\citenamefont{Liu et~al.}(2014{\natexlab{a}})\citenamefont{Liu, Zhou,
  Zhang, Wang, Weng, Prabhakaran, Mo, Shen, Fang, Dai
  et~al.}}]{Liu2014-Science}
\bibinfo{author}{\bibfnamefont{Z.~K.} \bibnamefont{Liu}},
  \bibinfo{author}{\bibfnamefont{B.}~\bibnamefont{Zhou}},
  \bibinfo{author}{\bibfnamefont{Y.}~\bibnamefont{Zhang}},
  \bibinfo{author}{\bibfnamefont{Z.~J.} \bibnamefont{Wang}},
  \bibinfo{author}{\bibfnamefont{H.~M.} \bibnamefont{Weng}},
  \bibinfo{author}{\bibfnamefont{D.}~\bibnamefont{Prabhakaran}},
  \bibinfo{author}{\bibfnamefont{S.-K.} \bibnamefont{Mo}},
  \bibinfo{author}{\bibfnamefont{Z.~X.} \bibnamefont{Shen}},
  \bibinfo{author}{\bibfnamefont{Z.}~\bibnamefont{Fang}},
  \bibinfo{author}{\bibfnamefont{X.}~\bibnamefont{Dai}}, \bibnamefont{et~al.},
  \bibinfo{journal}{Science} \textbf{\bibinfo{volume}{343}},
  \bibinfo{pages}{864} (\bibinfo{year}{2014}{\natexlab{a}}),
  \eprint{http://www.sciencemag.org/content/343/6173/864.full.pdf},
  \urlprefix\url{http://www.sciencemag.org/content/343/6173/864.abstract}.

\bibitem[{\citenamefont{Liu et~al.}(2014{\natexlab{b}})\citenamefont{Liu,
  Jiang, Zhou, Wang, Zhang, Weng, Prabhakaran, Mo, Peng, Dudin
  et~al.}}]{Liu2014-NatMater}
\bibinfo{author}{\bibfnamefont{Z.~K.} \bibnamefont{Liu}},
  \bibinfo{author}{\bibfnamefont{J.}~\bibnamefont{Jiang}},
  \bibinfo{author}{\bibfnamefont{B.}~\bibnamefont{Zhou}},
  \bibinfo{author}{\bibfnamefont{Z.~J.} \bibnamefont{Wang}},
  \bibinfo{author}{\bibfnamefont{Y.}~\bibnamefont{Zhang}},
  \bibinfo{author}{\bibfnamefont{H.~M.} \bibnamefont{Weng}},
  \bibinfo{author}{\bibfnamefont{D.}~\bibnamefont{Prabhakaran}},
  \bibinfo{author}{\bibfnamefont{S.-K.} \bibnamefont{Mo}},
  \bibinfo{author}{\bibfnamefont{H.}~\bibnamefont{Peng}},
  \bibinfo{author}{\bibfnamefont{P.}~\bibnamefont{Dudin}},
  \bibnamefont{et~al.}, \bibinfo{journal}{Nat. Mater.}
  \textbf{\bibinfo{volume}{13}}, \bibinfo{pages}{677}
  (\bibinfo{year}{2014}{\natexlab{b}}),
  \urlprefix\url{http://dx.doi.org/10.1038/nmat3990}.

\bibitem[{\citenamefont{Borisenko et~al.}(2014)\citenamefont{Borisenko, Gibson,
  Evtushinsky, Zabolotnyy, B\"uchner, and Cava}}]{Borisenko2014}
\bibinfo{author}{\bibfnamefont{S.}~\bibnamefont{Borisenko}},
  \bibinfo{author}{\bibfnamefont{Q.}~\bibnamefont{Gibson}},
  \bibinfo{author}{\bibfnamefont{D.}~\bibnamefont{Evtushinsky}},
  \bibinfo{author}{\bibfnamefont{V.}~\bibnamefont{Zabolotnyy}},
  \bibinfo{author}{\bibfnamefont{B.}~\bibnamefont{B\"uchner}},
  \bibnamefont{and} \bibinfo{author}{\bibfnamefont{R.~J.} \bibnamefont{Cava}},
  \bibinfo{journal}{Phys. Rev. Lett.} \textbf{\bibinfo{volume}{113}},
  \bibinfo{pages}{027603} (\bibinfo{year}{2014}),
  \urlprefix\url{http://link.aps.org/doi/10.1103/PhysRevLett.113.027603}.

\bibitem[{\citenamefont{Neupane et~al.}(2014)\citenamefont{Neupane, Xu, Sankar,
  Alidoust, Bian, Liu, Belopolski, Chang, Jeng, Lin et~al.}}]{Neupane2014}
\bibinfo{author}{\bibfnamefont{M.}~\bibnamefont{Neupane}},
  \bibinfo{author}{\bibfnamefont{S.-Y.} \bibnamefont{Xu}},
  \bibinfo{author}{\bibfnamefont{R.}~\bibnamefont{Sankar}},
  \bibinfo{author}{\bibfnamefont{N.}~\bibnamefont{Alidoust}},
  \bibinfo{author}{\bibfnamefont{G.}~\bibnamefont{Bian}},
  \bibinfo{author}{\bibfnamefont{C.}~\bibnamefont{Liu}},
  \bibinfo{author}{\bibfnamefont{I.}~\bibnamefont{Belopolski}},
  \bibinfo{author}{\bibfnamefont{T.-R.} \bibnamefont{Chang}},
  \bibinfo{author}{\bibfnamefont{H.-T.} \bibnamefont{Jeng}},
  \bibinfo{author}{\bibfnamefont{H.}~\bibnamefont{Lin}}, \bibnamefont{et~al.},
  \bibinfo{journal}{Nat. Commun.} \textbf{\bibinfo{volume}{5}},
  \bibinfo{pages}{3786} (\bibinfo{year}{2014}),
  \urlprefix\url{http://dx.doi.org/10.1038/ncomms4786}.

\bibitem[{\citenamefont{Young et~al.}(2012)\citenamefont{Young, Zaheer, Teo,
  Kane, Mele, and Rappe}}]{Young2012}
\bibinfo{author}{\bibfnamefont{S.~M.} \bibnamefont{Young}},
  \bibinfo{author}{\bibfnamefont{S.}~\bibnamefont{Zaheer}},
  \bibinfo{author}{\bibfnamefont{J.~C.~Y.} \bibnamefont{Teo}},
  \bibinfo{author}{\bibfnamefont{C.~L.} \bibnamefont{Kane}},
  \bibinfo{author}{\bibfnamefont{E.~J.} \bibnamefont{Mele}}, \bibnamefont{and}
  \bibinfo{author}{\bibfnamefont{A.~M.} \bibnamefont{Rappe}},
  \bibinfo{journal}{Phys. Rev. Lett.} \textbf{\bibinfo{volume}{108}},
  \bibinfo{pages}{140405} (\bibinfo{year}{2012}),
  \urlprefix\url{http://link.aps.org/doi/10.1103/PhysRevLett.108.140405}.

\bibitem[{\citenamefont{Kharzeev}(2014)}]{Kharzeev2014}
\bibinfo{author}{\bibfnamefont{D.~E.} \bibnamefont{Kharzeev}},
  \bibinfo{journal}{Prog. Part. Nucl. Phys.} \textbf{\bibinfo{volume}{75}},
  \bibinfo{pages}{133} (\bibinfo{year}{2014}),
  \urlprefix\url{http://www.sciencedirect.com/science/article/pii/S0146641014000039}.

\bibitem[{\citenamefont{He et~al.}(2014)\citenamefont{He, Hong, Dong, Pan,
  Zhang, Zhang, and Li}}]{He2014a}
\bibinfo{author}{\bibfnamefont{L.~P.} \bibnamefont{He}},
  \bibinfo{author}{\bibfnamefont{X.~C.} \bibnamefont{Hong}},
  \bibinfo{author}{\bibfnamefont{J.~K.} \bibnamefont{Dong}},
  \bibinfo{author}{\bibfnamefont{J.}~\bibnamefont{Pan}},
  \bibinfo{author}{\bibfnamefont{Z.}~\bibnamefont{Zhang}},
  \bibinfo{author}{\bibfnamefont{J.}~\bibnamefont{Zhang}}, \bibnamefont{and}
  \bibinfo{author}{\bibfnamefont{S.~Y.} \bibnamefont{Li}},
  \bibinfo{journal}{Phys. Rev. Lett.} \textbf{\bibinfo{volume}{113}},
  \bibinfo{pages}{246402} (\bibinfo{year}{2014}),
  \urlprefix\url{http://link.aps.org/doi/10.1103/PhysRevLett.113.246402}.

\bibitem[{\citenamefont{Liang et~al.}(2015)\citenamefont{Liang, Gibson, Ali,
  Liu, Cava, and Ong}}]{Liang2015c}
\bibinfo{author}{\bibfnamefont{T.}~\bibnamefont{Liang}},
  \bibinfo{author}{\bibfnamefont{Q.}~\bibnamefont{Gibson}},
  \bibinfo{author}{\bibfnamefont{M.~N.} \bibnamefont{Ali}},
  \bibinfo{author}{\bibfnamefont{M.}~\bibnamefont{Liu}},
  \bibinfo{author}{\bibfnamefont{R.~J.} \bibnamefont{Cava}}, \bibnamefont{and}
  \bibinfo{author}{\bibfnamefont{N.~P.} \bibnamefont{Ong}},
  \bibinfo{journal}{Nat. Mater.} \textbf{\bibinfo{volume}{14}},
  \bibinfo{pages}{280} (\bibinfo{year}{2015}),
  \urlprefix\url{http://dx.doi.org/10.1038/nmat4143}.

\bibitem[{\citenamefont{Xu et~al.}(2015)\citenamefont{Xu, Liu, Kushwaha,
  Sankar, Krizan, Belopolski, Neupane, Bian, Alidoust, Chang
  et~al.}}]{Xu2015-Science}
\bibinfo{author}{\bibfnamefont{S.-Y.} \bibnamefont{Xu}},
  \bibinfo{author}{\bibfnamefont{C.}~\bibnamefont{Liu}},
  \bibinfo{author}{\bibfnamefont{S.~K.} \bibnamefont{Kushwaha}},
  \bibinfo{author}{\bibfnamefont{R.}~\bibnamefont{Sankar}},
  \bibinfo{author}{\bibfnamefont{J.~W.} \bibnamefont{Krizan}},
  \bibinfo{author}{\bibfnamefont{I.}~\bibnamefont{Belopolski}},
  \bibinfo{author}{\bibfnamefont{M.}~\bibnamefont{Neupane}},
  \bibinfo{author}{\bibfnamefont{G.}~\bibnamefont{Bian}},
  \bibinfo{author}{\bibfnamefont{N.}~\bibnamefont{Alidoust}},
  \bibinfo{author}{\bibfnamefont{T.-R.} \bibnamefont{Chang}},
  \bibnamefont{et~al.}, \bibinfo{journal}{Science}
  \textbf{\bibinfo{volume}{347}}, \bibinfo{pages}{294} (\bibinfo{year}{2015}),
  \eprint{http://www.sciencemag.org/content/347/6219/294.full.pdf},
  \urlprefix\url{http://www.sciencemag.org/content/347/6219/294.abstract}.

\bibitem[{\citenamefont{Chen et~al.}(2013)\citenamefont{Chen, Wu, and
  Burkov}}]{Chen2013a}
\bibinfo{author}{\bibfnamefont{Y.}~\bibnamefont{Chen}},
  \bibinfo{author}{\bibfnamefont{S.}~\bibnamefont{Wu}}, \bibnamefont{and}
  \bibinfo{author}{\bibfnamefont{A.~A.} \bibnamefont{Burkov}},
  \bibinfo{journal}{Phys. Rev. B} \textbf{\bibinfo{volume}{88}},
  \bibinfo{pages}{125105} (\bibinfo{year}{2013}),
  \urlprefix\url{http://link.aps.org/doi/10.1103/PhysRevB.88.125105}.

\bibitem[{\citenamefont{Hosur and Qi}(2015)}]{Hosur2014}
\bibinfo{author}{\bibfnamefont{P.}~\bibnamefont{Hosur}} \bibnamefont{and}
  \bibinfo{author}{\bibfnamefont{X.-L.} \bibnamefont{Qi}},
  \bibinfo{journal}{Phys. Rev. B} \textbf{\bibinfo{volume}{91}},
  \bibinfo{pages}{081106} (\bibinfo{year}{2015}),
  \urlprefix\url{http://link.aps.org/doi/10.1103/PhysRevB.91.081106}.

\bibitem[{\citenamefont{Potter et~al.}(2014)\citenamefont{Potter, Kimchi, and
  Vishwanath}}]{Potter2014}
\bibinfo{author}{\bibfnamefont{A.~C.} \bibnamefont{Potter}},
  \bibinfo{author}{\bibfnamefont{I.}~\bibnamefont{Kimchi}}, \bibnamefont{and}
  \bibinfo{author}{\bibfnamefont{A.}~\bibnamefont{Vishwanath}},
  \bibinfo{journal}{Nat. Commun.} \textbf{\bibinfo{volume}{5}},
  \bibinfo{pages}{5161} (\bibinfo{year}{2014}),
  \urlprefix\url{http://dx.doi.org/10.1038/ncomms6161}.

\bibitem[{\citenamefont{Burkov and Balents}(2011)}]{PhysRevLett.107.127205}
\bibinfo{author}{\bibfnamefont{A.~A.} \bibnamefont{Burkov}} \bibnamefont{and}
  \bibinfo{author}{\bibfnamefont{L.}~\bibnamefont{Balents}},
  \bibinfo{journal}{Phys. Rev. Lett.} \textbf{\bibinfo{volume}{107}},
  \bibinfo{pages}{127205} (\bibinfo{year}{2011}),
  \urlprefix\url{http://link.aps.org/doi/10.1103/PhysRevLett.107.127205}.

\bibitem[{\citenamefont{Burkov et~al.}(2011)\citenamefont{Burkov, Hook, and
  Balents}}]{PhysRevB.84.235126}
\bibinfo{author}{\bibfnamefont{A.~A.} \bibnamefont{Burkov}},
  \bibinfo{author}{\bibfnamefont{M.~D.} \bibnamefont{Hook}}, \bibnamefont{and}
  \bibinfo{author}{\bibfnamefont{L.}~\bibnamefont{Balents}},
  \bibinfo{journal}{Phys. Rev. B} \textbf{\bibinfo{volume}{84}},
  \bibinfo{pages}{235126} (\bibinfo{year}{2011}),
  \urlprefix\url{http://link.aps.org/doi/10.1103/PhysRevB.84.235126}.

\bibitem[{\citenamefont{Gorbar et~al.}(2013)\citenamefont{Gorbar, Miransky, and
  Shovkovy}}]{Gorbar-MagneticDiracWeyl}
\bibinfo{author}{\bibfnamefont{E.~V.} \bibnamefont{Gorbar}},
  \bibinfo{author}{\bibfnamefont{V.~A.} \bibnamefont{Miransky}},
  \bibnamefont{and} \bibinfo{author}{\bibfnamefont{I.~A.}
  \bibnamefont{Shovkovy}}, \bibinfo{journal}{Phys. Rev. B}
  \textbf{\bibinfo{volume}{88}}, \bibinfo{pages}{165105}
  (\bibinfo{year}{2013}),
  \urlprefix\url{http://link.aps.org/doi/10.1103/PhysRevB.88.165105}.

\bibitem[{\citenamefont{Phillips and Aji}(2014)}]{PhysRevB.90.115111}
\bibinfo{author}{\bibfnamefont{M.}~\bibnamefont{Phillips}} \bibnamefont{and}
  \bibinfo{author}{\bibfnamefont{V.}~\bibnamefont{Aji}},
  \bibinfo{journal}{Phys. Rev. B} \textbf{\bibinfo{volume}{90}},
  \bibinfo{pages}{115111} (\bibinfo{year}{2014}),
  \urlprefix\url{http://link.aps.org/doi/10.1103/PhysRevB.90.115111}.

\bibitem[{\citenamefont{Narayanan et~al.}(2015)\citenamefont{Narayanan, Watson,
  Blake, Bruyant, Drigo, Chen, Prabhakaran, Yan, Felser, Kong
  et~al.}}]{PhysRevLett.114.117201}
\bibinfo{author}{\bibfnamefont{A.}~\bibnamefont{Narayanan}},
  \bibinfo{author}{\bibfnamefont{M.~D.} \bibnamefont{Watson}},
  \bibinfo{author}{\bibfnamefont{S.~F.} \bibnamefont{Blake}},
  \bibinfo{author}{\bibfnamefont{N.}~\bibnamefont{Bruyant}},
  \bibinfo{author}{\bibfnamefont{L.}~\bibnamefont{Drigo}},
  \bibinfo{author}{\bibfnamefont{Y.~L.} \bibnamefont{Chen}},
  \bibinfo{author}{\bibfnamefont{D.}~\bibnamefont{Prabhakaran}},
  \bibinfo{author}{\bibfnamefont{B.}~\bibnamefont{Yan}},
  \bibinfo{author}{\bibfnamefont{C.}~\bibnamefont{Felser}},
  \bibinfo{author}{\bibfnamefont{T.}~\bibnamefont{Kong}}, \bibnamefont{et~al.},
  \bibinfo{journal}{Phys. Rev. Lett.} \textbf{\bibinfo{volume}{114}},
  \bibinfo{pages}{117201} (\bibinfo{year}{2015}),
  \urlprefix\url{http://link.aps.org/doi/10.1103/PhysRevLett.114.117201}.

\bibitem[{\citenamefont{Cao et~al.}(2014)\citenamefont{Cao, Liang, Zhang, Liu,
  Huang, Jin, Chen, Wang, Wang, Zhao et~al.}}]{JCao}
\bibinfo{author}{\bibfnamefont{J.}~\bibnamefont{Cao}},
  \bibinfo{author}{\bibfnamefont{S.}~\bibnamefont{Liang}},
  \bibinfo{author}{\bibfnamefont{C.}~\bibnamefont{Zhang}},
  \bibinfo{author}{\bibfnamefont{Y.}~\bibnamefont{Liu}},
  \bibinfo{author}{\bibfnamefont{J.}~\bibnamefont{Huang}},
  \bibinfo{author}{\bibfnamefont{Z.}~\bibnamefont{Jin}},
  \bibinfo{author}{\bibfnamefont{Z.-G.} \bibnamefont{Chen}},
  \bibinfo{author}{\bibfnamefont{Z.}~\bibnamefont{Wang}},
  \bibinfo{author}{\bibfnamefont{Q.}~\bibnamefont{Wang}},
  \bibinfo{author}{\bibfnamefont{J.}~\bibnamefont{Zhao}}, \bibnamefont{et~al.}
  (\bibinfo{year}{2014}), \bibinfo{note}{arXiv:1412.0824 (unpublished)},
  \urlprefix\url{http://arxiv.org/abs/1412.0824}.

\bibitem[{\citenamefont{Weng et~al.}(2014)\citenamefont{Weng, Dai, and
  Fang}}]{PhysRevX.4.011002}
\bibinfo{author}{\bibfnamefont{H.}~\bibnamefont{Weng}},
  \bibinfo{author}{\bibfnamefont{X.}~\bibnamefont{Dai}}, \bibnamefont{and}
  \bibinfo{author}{\bibfnamefont{Z.}~\bibnamefont{Fang}},
  \bibinfo{journal}{Phys. Rev. X} \textbf{\bibinfo{volume}{4}},
  \bibinfo{pages}{011002} (\bibinfo{year}{2014}),
  \urlprefix\url{http://link.aps.org/doi/10.1103/PhysRevX.4.011002}.

\bibitem[{\citenamefont{Li et~al.}(2014)\citenamefont{Li, Kharzeev, Zhang,
  Huang, Pletikosic, Fedorov, Zhong, Schneeloch, Gu, and Valla}}]{LiQ}
\bibinfo{author}{\bibfnamefont{Q.}~\bibnamefont{Li}},
  \bibinfo{author}{\bibfnamefont{D.~E.} \bibnamefont{Kharzeev}},
  \bibinfo{author}{\bibfnamefont{C.}~\bibnamefont{Zhang}},
  \bibinfo{author}{\bibfnamefont{Y.}~\bibnamefont{Huang}},
  \bibinfo{author}{\bibfnamefont{I.}~\bibnamefont{Pletikosic}},
  \bibinfo{author}{\bibfnamefont{A.~V.} \bibnamefont{Fedorov}},
  \bibinfo{author}{\bibfnamefont{R.~D.} \bibnamefont{Zhong}},
  \bibinfo{author}{\bibfnamefont{J.~A.} \bibnamefont{Schneeloch}},
  \bibinfo{author}{\bibfnamefont{G.~D.} \bibnamefont{Gu}}, \bibnamefont{and}
  \bibinfo{author}{\bibfnamefont{T.}~\bibnamefont{Valla}}
  (\bibinfo{year}{2014}), \bibinfo{note}{arXiv:1412.6543},
  \urlprefix\url{http://arxiv.org/abs/1412.6543}.

\bibitem[{\citenamefont{Chen et~al.}(2015)\citenamefont{Chen, Zhang,
  Schneeloch, Zhang, Li, Gu, and Wang}}]{RYChen}
\bibinfo{author}{\bibfnamefont{R.~Y.} \bibnamefont{Chen}},
  \bibinfo{author}{\bibfnamefont{S.~J.} \bibnamefont{Zhang}},
  \bibinfo{author}{\bibfnamefont{J.~A.} \bibnamefont{Schneeloch}},
  \bibinfo{author}{\bibfnamefont{C.}~\bibnamefont{Zhang}},
  \bibinfo{author}{\bibfnamefont{Q.}~\bibnamefont{Li}},
  \bibinfo{author}{\bibfnamefont{G.~D.} \bibnamefont{Gu}}, \bibnamefont{and}
  \bibinfo{author}{\bibfnamefont{N.~L.} \bibnamefont{Wang}},
  \bibinfo{journal}{Phys. Rev. B} \textbf{\bibinfo{volume}{92}},
  \bibinfo{pages}{075107} (\bibinfo{year}{2015}),
  \urlprefix\url{http://link.aps.org/doi/10.1103/PhysRevB.92.075107}.

\bibitem[{\citenamefont{Hosur et~al.}(2012)\citenamefont{Hosur, Parameswaran,
  and Vishwanath}}]{Hosur2012}
\bibinfo{author}{\bibfnamefont{P.}~\bibnamefont{Hosur}},
  \bibinfo{author}{\bibfnamefont{S.~A.} \bibnamefont{Parameswaran}},
  \bibnamefont{and}
  \bibinfo{author}{\bibfnamefont{A.}~\bibnamefont{Vishwanath}},
  \bibinfo{journal}{Phys. Rev. Lett.} \textbf{\bibinfo{volume}{108}},
  \bibinfo{pages}{046602} (\bibinfo{year}{2012}),
  \urlprefix\url{http://link.aps.org/doi/10.1103/PhysRevLett.108.046602}.

\bibitem[{\citenamefont{Ashby and Carbotte}(2013)}]{Ashby2013}
\bibinfo{author}{\bibfnamefont{P.~E.~C.} \bibnamefont{Ashby}} \bibnamefont{and}
  \bibinfo{author}{\bibfnamefont{J.~P.} \bibnamefont{Carbotte}},
  \bibinfo{journal}{Phys. Rev. B} \textbf{\bibinfo{volume}{87}},
  \bibinfo{pages}{245131} (\bibinfo{year}{2013}),
  \urlprefix\url{http://link.aps.org/doi/10.1103/PhysRevB.87.245131}.

\bibitem[{SI()}]{SI}
\bibinfo{note}{Supplementary information}.

\bibitem[{\citenamefont{Orlita et~al.}(2015)\citenamefont{Orlita, Piot,
  Martinez, Kumar, Faugeras, Potemski, Michel, Hankiewicz, Brauner,
  Dra\ifmmode~\check{s}\else \v{s}\fi{}ar et~al.}}]{Bi2Se3}
\bibinfo{author}{\bibfnamefont{M.}~\bibnamefont{Orlita}},
  \bibinfo{author}{\bibfnamefont{B.~A.} \bibnamefont{Piot}},
  \bibinfo{author}{\bibfnamefont{G.}~\bibnamefont{Martinez}},
  \bibinfo{author}{\bibfnamefont{N.~K.~S.} \bibnamefont{Kumar}},
  \bibinfo{author}{\bibfnamefont{C.}~\bibnamefont{Faugeras}},
  \bibinfo{author}{\bibfnamefont{M.}~\bibnamefont{Potemski}},
  \bibinfo{author}{\bibfnamefont{C.}~\bibnamefont{Michel}},
  \bibinfo{author}{\bibfnamefont{E.~M.} \bibnamefont{Hankiewicz}},
  \bibinfo{author}{\bibfnamefont{T.}~\bibnamefont{Brauner}},
  \bibinfo{author}{\bibfnamefont{i.~c.~v.}
  \bibnamefont{Dra\ifmmode~\check{s}\else \v{s}\fi{}ar}}, \bibnamefont{et~al.},
  \bibinfo{journal}{Phys. Rev. Lett.} \textbf{\bibinfo{volume}{114}},
  \bibinfo{pages}{186401} (\bibinfo{year}{2015}),
  \urlprefix\url{http://link.aps.org/doi/10.1103/PhysRevLett.114.186401}.

\bibitem[{\citenamefont{Sadowski et~al.}(2006)\citenamefont{Sadowski, Martinez,
  Potemski, Berger, and de~Heer}}]{PhysRevLett.97.266405}
\bibinfo{author}{\bibfnamefont{M.~L.} \bibnamefont{Sadowski}},
  \bibinfo{author}{\bibfnamefont{G.}~\bibnamefont{Martinez}},
  \bibinfo{author}{\bibfnamefont{M.}~\bibnamefont{Potemski}},
  \bibinfo{author}{\bibfnamefont{C.}~\bibnamefont{Berger}}, \bibnamefont{and}
  \bibinfo{author}{\bibfnamefont{W.~A.} \bibnamefont{de~Heer}},
  \bibinfo{journal}{Phys. Rev. Lett.} \textbf{\bibinfo{volume}{97}},
  \bibinfo{pages}{266405} (\bibinfo{year}{2006}),
  \urlprefix\url{http://link.aps.org/doi/10.1103/PhysRevLett.97.266405}.

\bibitem[{\citenamefont{Jeon et~al.}(2014)\citenamefont{Jeon, Zhou, Gyenis,
  Feldman, Kimchi, Potter, Gibson, Cava, Vishwanath, and Yazdani}}]{Jeon2014}
\bibinfo{author}{\bibfnamefont{S.}~\bibnamefont{Jeon}},
  \bibinfo{author}{\bibfnamefont{B.~B.} \bibnamefont{Zhou}},
  \bibinfo{author}{\bibfnamefont{A.}~\bibnamefont{Gyenis}},
  \bibinfo{author}{\bibfnamefont{B.~E.} \bibnamefont{Feldman}},
  \bibinfo{author}{\bibfnamefont{I.}~\bibnamefont{Kimchi}},
  \bibinfo{author}{\bibfnamefont{A.~C.} \bibnamefont{Potter}},
  \bibinfo{author}{\bibfnamefont{Q.~D.} \bibnamefont{Gibson}},
  \bibinfo{author}{\bibfnamefont{R.~J.} \bibnamefont{Cava}},
  \bibinfo{author}{\bibfnamefont{A.}~\bibnamefont{Vishwanath}},
  \bibnamefont{and} \bibinfo{author}{\bibfnamefont{A.}~\bibnamefont{Yazdani}},
  \bibinfo{journal}{Nat. Mater.} \textbf{\bibinfo{volume}{13}},
  \bibinfo{pages}{851 } (\bibinfo{year}{2014}),
  \urlprefix\url{http://dx.doi.org/10.1038/nmat4023}.

\end{thebibliography}

\end{document}